\documentclass[aps,twocolumn,amsmath,letterpaper]{revtex4}
\usepackage{amssymb}
\usepackage{graphicx}
\usepackage{array}
\usepackage{hhline}
\usepackage{longtable}

\renewcommand{\thetable}{\Roman{table}} \thetable

\begin{document}

\title{Multicritical Point Relations in
\\Three Dual Pairs of Hierarchical-Lattice Ising Spin-Glasses}
\author{Michael Hinczewski and A. Nihat Berker}
\affiliation{Department of Physics, Istanbul Technical University,
Maslak 34469, Istanbul, Turkey,} \affiliation{Department of Physics,
Massachusetts Institute of Technology, Cambridge, Massachusetts
02139, U.S.A.,} \affiliation{Feza G\"ursey Research Institute,
T\"UBITAK - Bosphorus University, \c{C}engelk\"oy 34680, Istanbul,
Turkey}
\begin{abstract}  The Ising spin-glasses are investigated on
three dual pairs of hierarchical lattices, using exact
renormalization-group transformation of the quenched bond
probability distribution.  The goal is to investigate a recent
conjecture which relates, on such pairs of dual lattices, the
locations of the multicritical points, which occur on the Nishimori
symmetry line. Towards this end we precisely determine the global
phase diagrams for these six hierarchical spin-glasses, using up to
$2.5 \times 10^9$ probability bins to represent the quenched
distribution subjected to an exact renormalization-group
transformation.  We find in all three cases that the conjecture is
realized to a very good approximation, even when the mutually dual
models belong to different spatial dimensionalities $d$ and have
different phase diagram topologies at the multicritical points of
the conjecture and even though the contributions to the conjecture
from each lattice of the dual pair are strongly asymmetric.  In all
six phase diagrams, we find reentrance near the multicritical point.
In the models with $d=2$ or 1.5, the spin-glass phase does not occur
and the phase boundary between the ferromagnetic and paramagnetic
phases is second order with a strong violation of universality.

PACS numbers: 75.10.Nr, 64.60.Kw, 05.45.Df, 05.10.Cc
\end{abstract}
\maketitle
\def\s{\rule{0in}{0.28in}}

\section{Introduction}
\setlength{\LTcapwidth}{\columnwidth}

The phase diagram structure of spin-glasses remains an open field of
inquiry, since most approaches to the problem rely on
approximations.  Any exact analytical result in this area is thus
very valuable, both for the direct information it provides and as a
test for approximation methods.  Over the last few years striking
progress has been made combining the replica method, duality, and
symmetry arguments~\cite{Nishimori, NishimoriNemoto, Maillard,
Takeda2, Takeda1}, an approach which has yielded the exact locations
of the multicritical points in the Ising and Potts spin-glasses on
the square lattice and in the four-dimensional random-plaquette
gauge model.  The most recent result in this series~\cite{Takeda1}
is a general conjecture relating the multicritical point locations
of any spin-glasses on a pair of mutually dual lattices.  In support
of the conjecture, estimates based on Monte Carlo simulations were
given for Ising spin-glasses, in $d=2$, on the dual pairs of
triangular and hexagonal lattices and, in $d=3$, on the dual pairs
of bilinear and lattice-gauge interactions on the cubic lattice.  In
both cases, within the numerical limitations, the conjecture is
approximately satisfied.

We propose here to extensively test the conjecture in an alternative
fashion using hierarchical lattices~\cite{BerkerOstlund, Kaufman,
Kaufman2}, by looking at Ising spin-glasses on mutually dual pairs
\cite{Kaufman3, Kaufman4, Itzykson, Ottavi} of such lattices. These
constitute ideal testing grounds, since an exact
renormalization-group transformation for the quenched bond
probability distribution can be constructed for such lattices,
yielding global phase diagrams and critical properties. Accordingly,
the location of the phase boundaries and of the multicritical points
are precisely determined. We thus investigate three pairs of
hierarchical lattices, and in the end find that the conjecture is
very nearly satisfied for all of them.

\section{The Conjecture}

The Ising spin-glass is given by the Hamiltonian
\begin{equation}\label{eq:1}
-\beta H = \sum_{\langle ij \rangle} J_{ij} s_i s_j\,,
\end{equation}
where $s_i = \pm 1$ at each site $i$, $\langle i j \rangle$ denotes
a sum over nearest-neighbor pairs of sites, and the bond strengths
$J_{ij}$ are equal to $+J$ with probability $1-p$ and $-J$ with
probability $p$.  The limits $p=0$ and $p=1$ correspond to purely
ferromagnetic and purely antiferromagnetic systems respectively.

To give a brief overview of the conjecture, let us consider the
model on an arbitrary lattice, and treat the randomness through the
replica method, where the system is replicated $n$ times and the $n
\to 0$ limit is eventually taken, in order to get results for the
physical system.  The partition function of the $n$-replicated
system after averaging over randomness, $Z_n$, can be expressed
entirely as a function of $n+1$ ``averaged'' combinations of edge
Boltzmann factors, $e^{\pm J}$, associated with nearest-neighbor
bonds~\cite{Maillard, Takeda2}. These averaged Boltzmann factors,
$x_k(p,J)$, $k=0,\ldots,n$, have the form
\begin{equation}\label{eq:1a}
x_k(p,J) = p e^{-(n-k)J}e^{k J} + (1-p) e^{(n-k)J} e^{-kJ}\,,
\end{equation}
where the $k$th factor corresponds to a configuration with a
parallel-spin bond in $n-k$ replicas and an antiparallel-spin bond in
$k$ replicas~\cite{Takeda1}.  Thus,
\begin{equation}\label{eq:1b}
Z_n = Z_n(x_0(p,J),\,x_1(p,J),\,\ldots,\,x_n(p,J))\,.
\end{equation}

The partition function on the dual lattice, $Z^\ast_n$, can be expressed in a
similar form,
\begin{equation}\label{eq:1c}
Z^\ast_n = Z^\ast_n (x^\ast_0(p,J),\,x^\ast_1(p,J),\,\ldots,\,x^\ast_n(p,J))\,,
\end{equation}
with the dual counterparts to the averaged Boltzmann factors given
by
\begin{equation}\label{eq:1d}
\begin{split}
x^\ast_{2k}(p,J) &= \left(\frac{e^{-J} + e^{J}}{\sqrt{2}}\right)^{n-2k} \left(\frac{e^{-J} - e^{J}}{\sqrt{2}}\right)^{2k}\,,\\
x^\ast_{2k+1}(p,J) &= (2p-1)\left(\frac{e^{-J} +
  e^{J}}{\sqrt{2}}\right)^{n-2k-1}\\
& \qquad \cdot \left(\frac{e^{-J} -
e^{J}}{\sqrt{2}}\right)^{2k+1}\,,
\end{split}
\end{equation}
for $0 \le 2k < 2k+1 \le n$.  $Z_n$ and $Z_n^\ast$ are related as
\cite{Takeda1}
\begin{multline}\label{eq:1e}
Z_n(x_0(p,J),\,\ldots,\,x_n(p,J))\\
= 2^a Z_n^\ast(x_0^\ast(p,J),\,\ldots,\,x_n^\ast(p,J))\,,
\end{multline}
where $a$ is a constant, which can be eliminated by using
Eq.~\eqref{eq:1e} evaluated at two different sets of parameters,
$(p_1,J_1)$ and $(p_2,J_2)$, giving a relationship of the form
\begin{equation}\label{eq:1f}
\begin{split}
&Z_n(x_0(p_1,J_1),\,\ldots,\,x_n(p_1,J_1))\\
&\qquad \cdot Z_n^\ast(x_0(p_2,J_2),\,\ldots,\,x_n(p_2,J_2))\\
&= Z_n^\ast(x^\ast_0(p_1,J_1),\,\ldots,\,x^\ast_n(p_1,J_1))\\
&\qquad \cdot Z_n(x^\ast_0(p_2,J_2),\,\ldots,\,x^\ast_n(p_2,J_2))\,.
\end{split}
\end{equation}
The individual partition functions $Z_n$ can be rewritten by
extracting $x_0$, the averaged Boltzmann factor corresponding to an
all-parallel spin state, thus effectively measuring the energy of
the system relative to this state~\cite{Maillard}:
\begin{equation}\label{eq:1g}
Z_n(x_0,x_1,\ldots,x_n) = x_0^{N_B} {\cal Z}_n(u_1, u_2, \ldots,
u_n)\,,
\end{equation}
where $N_B$ is the number of bonds in the lattice, and the reduced
variables are $u_i \equiv x_i / x_0$.  Eq.~\eqref{eq:1f} becomes
\begin{equation}\label{eq:1h}
\begin{split}
&[x_0(p_1,J_1)x_0(p_2,J_2)]^{N_B}
{\cal Z}_n(u_1(p_1,J_1),\,\ldots,\,u_n(p_1,J_1))\\
&\qquad \cdot {\cal Z}_n^\ast(u_1(p_2,J_2),\,\ldots,\,u_n(p_2,J_2))\\
&=[x^\ast_0(p_1,J_1)x^\ast_0(p_2,J_2) ]^{N_B}
{\cal Z}_n^\ast(u^\ast_1(p_1,J_1),\,\ldots,\,u^\ast_n(p_1,J_1))\\
&\qquad \cdot {\cal
Z}_n(u^\ast_1(p_2,J_2),\,\ldots,\,u^\ast_n(p_2,J_2))\,.
\end{split}
\end{equation}

In general, the form of Eq.~\eqref{eq:1h} is too complicated to
yield useful information relating the locations of phase
transitions.  However, the multicritical points in both original and
dual systems are expected to lie~\cite{LeDoussal, LeDoussal2,
Hartford} on the Nishimori line~\cite{Nishimori}, which simplifies
the relation. Furthermore, the conjecture advanced in
Ref.~\cite{Takeda1} states that, for the multicritical points
$(p_{1m}, J_{1m})$ of the original system and $(p_{2m}, J_{2m})$ of
its dual, Eq.~\eqref{eq:1h} is satisfied when the leading Boltzmann
factors $x_0$ from each side are equal,
\begin{equation}\label{eq:1i}
x_0(p_{1m},J_{1m})x_0(p_{2m},J_{2m}) = x^\ast_0(p_{1m},J_{1m})
x^\ast_0(p_{2m},J_{2m})\,.
\end{equation}
Since $(p_{1m}, J_{1m})$ and $(p_{2m}, J_{2m})$ lie on the Nishimori
line,
\begin{equation}\label{eq:1j}
e^{2J_{1m}} = \frac{1-p_{1m}}{p_{1m}}\,, \qquad e^{2J_{2m}} =
\frac{1-p_{2m}}{p_{2m}}\,.
\end{equation}
From Eqs.~\eqref{eq:1a} and \eqref{eq:1d}, Eq.~\eqref{eq:1i} gives
\begin{equation}\label{eq:1k}
(p_{1m}^{n+1} + (1-p_{1m})^{n+1})(p_{2m}^{n+1} + (1-p_{2m})^{n+1}) =
2^{-n}\,.
\end{equation}
Finally taking the limit, $n \to 0$, one obtains the condition
\begin{equation}\label{eq:1l}
H(p_{1m}) + H(p_{2m}) = 1\,,
\end{equation}
where $H(p) \equiv -p \log_2 p - (1-p) \log_2 (1-p)$.  As expressed
in Eq.~\eqref{eq:1l}, the conjecture is asserted to hold for
multicritical points of Ising spin-glasses on any pair of mutually
dual lattices~\cite{Takeda1}.

\section{The Multitude of Ising Spin-Glasses on Hierarchical Lattices}

Hierarchical lattices~\cite{BerkerOstlund, Kaufman, Kaufman2} are
constructed by replacing every single bond, in a connected cluster
of bonds, with the connected cluster of bonds itself, and repeating
this step an infinite number of times.  These provide models exactly
solvable by renormalization group, with which complex problems have
been studied and understood. For example, frustrated \cite{McKay},
spin-glass \cite{Migliorini}, random-bond \cite{Andelman} and
random-field \cite{Falicov2}, Schr\"odinger equation \cite{Domany},
lattice-vibration \cite{Langlois}, dynamic scaling
\cite{Stinchcombe}, aperiodic magnet \cite{Haddad}, complex phase
diagram \cite{Le}, and directed-path \cite{daSilveira} systems,
etc., have been solved on hierarchical lattices.

To test the conjecture of Eq.~\eqref{eq:1l}, we study Ising
spin-glasses on the dual pairs of hierarchical lattices, depicted in
Figs.~\ref{lfig1}, \ref{lfig2}, and \ref{lfig3}.  Each lattice in a
given pair is the dual of the other. These particular choice of
lattices are motivated by their properties under
renormalization-group transformation as related to physical
lattices.  The hierarchical lattices of Fig.~1(a) and (b) yield the
two variants of the Migdal-Kadanoff recursion
relations~\cite{Migdal, Kadanoff} for dimension $d=2$ with length
rescaling factor $b=3$.  Similarly, the lattice in Fig.~2(a) yields
a Migdal-Kadanoff recursion relation for $d=3$, $b=3$.  Its dual
lattice in Fig.~2(b) has $d=1.5, b=9$.  (The two variants of the
Migdal-Kadanoff recursion relations correspond to mutually dual
hierarchical lattices only in $d=2$.)  Lastly, the hybrid lattice in
Fig.~3(a) is interesting because it has been shown to give very
accurate results for the critical temperatures of the $d=3$
isotropic and anisotropic Ising
model~\cite{ErbasTuncerYucesoyBerker}.  This lattice has $d=3, b=3$,
while its dual in Fig.~3(b) has $d=1.5, b=9$.

\section{Exact Renormalization-Group Transformation of Hierarchical Spin-Glasses}

\begin{figure}[h]
\includegraphics*[scale=0.5]{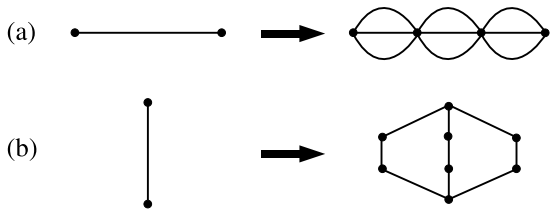}
\caption{The pair of mutually dual hierarchical lattices on which
the $d=2$, $b=3$ Migdal-Kadanoff recursion relations are
exact.}\label{lfig1}
\end{figure}
\begin{figure}[h]
\includegraphics*[scale=0.5]{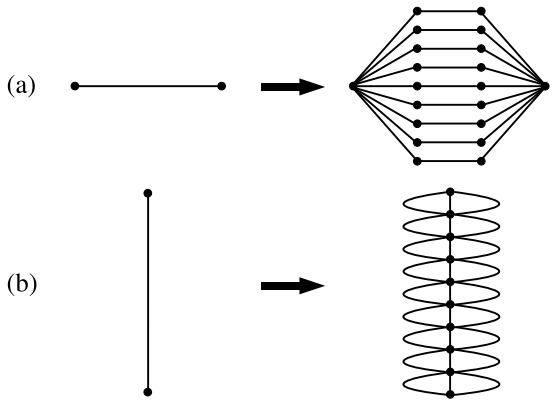}
\caption{Another pair of mutually dual hierarchical lattices.  The
Migdal-Kadanoff recursion relations are exact for lattice (a) with
$d=3$, $b=3$.  Its dual lattice, in (b), has $d=1.5$, $b=9$.}
\label{lfig2}
\end{figure}
\begin{figure}[h]
\includegraphics*[scale=0.5]{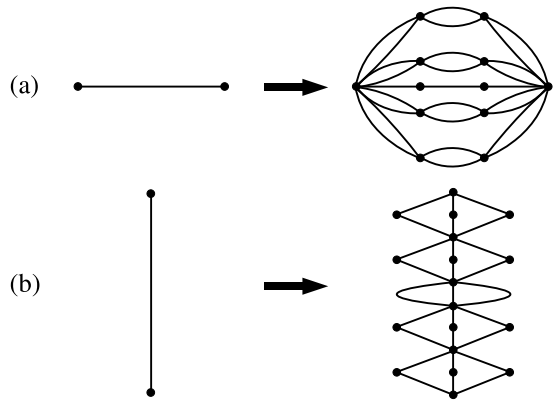}
\caption{Another pair of mutually dual hierarchical lattices, with
$d=3$, $b=3$ and $d=1.5$, $b=9$ respectively.} \label{lfig3}
\end{figure}

For a pure system, the renormalization-group transformation on a
hierarchical lattice consists of a decimation by summing over the
internal sites in each of the connected clusters making up the
lattice (the right-hand sides of Figs. \ref{lfig1}-\ref{lfig3}).
Thus, the hierarchical lattice construction process is reversed, as
each connected cluster is replaced by a single renormalized bond.
The decimation can be expressed as a mapping,
\begin{equation}\label{eq:2}
J^\prime_{i^\prime j^\prime} = R(\{J_{ij}\})\,,
\end{equation}
where the set $\{J_{ij}\}$ are all the bonds within the connected
cluster of the original system and $J^\prime_{i^\prime j^\prime}$ is
the renormalized bond between sites $i^\prime$ and $j^\prime$ of the
rescaled system.  In the pure case, all $J_{ij}$ bonds are
independent for $ij$, and the implementation of Eq.~\eqref{eq:2} is
straightforward.

When quenched randomness is added to the system, the
renormalization-group transformation is expressed in terms of
quenched probability distributions~\cite{Andelman}, where the
quenched probability distribution ${\cal
P}^\prime(J^\prime_{i^\prime j^\prime})$ in the rescaled system is
calculated from ${\cal P}(J_{ij})$ in the original system through
the convolution
\begin{equation}\label{eq:3}
{\cal P}^\prime(J^\prime_{i^\prime j^\prime}) = \int
\left[\prod_{ij}^{i^\prime j^\prime} dJ_{ij}\, {\cal
P}(J_{ij})\right] \delta\left(J^\prime_{i^\prime j^\prime} -
R(\{J_{ij}\})\right)\,.
\end{equation}
Here the product runs over all the bonds $ij$ in the connected
cluster of the original system between sites $i^\prime$ and
$j^\prime$.

The recursion of the quenched probability distribution,
Eq.~\eqref{eq:3}, is implemented numerically.  The probability
distribution is represented by histograms, each histogram being
specified by a bond strength and an associated probability. Thus,
for the spin-glass problem, the starting distribution consists of
two histograms, one at $J$ with probability $1-p$, and one at $-J$
with probability $p$. Eq.~\eqref{eq:3} dictates the convolution of 9
probability distributions for the lattices of Fig.~\ref{lfig1}, and
the convolution of 27 distributions for the lattices of
Figs.~\ref{lfig2} and \ref{lfig3}.  In this task, computational
storage limits can be maximally exploited by factorizing
Eq.~\eqref{eq:3} into an equivalent series of pairwise convolutions,
each of which involves only two distributions convoluted using an
appropriate $R$ function. The types of pairwise convolutions needed
are a ``bond-moving'' convolution, with
\begin{equation}\label{eq:4}
R_\text{bm}(J_{i_1 j_1},J_{i_2 j_2}) = J_{i_1 j_1}
+ J_{i_2 j_2}\,,
\end{equation}
and a decimation convolution, with
\begin{equation}\label{eq:5}
R_\text{dc}(J_{i_1 j_1},J_{i_2 j_2}) = \frac{1}{2}
\ln\left[\frac{\cosh(J_{i_1 j_1}+J_{i_2 j_2})}{\cosh(J_{i_1
      j_1}-J_{i_2 j_2})}
\right]\,,
\end{equation}
which is just the standard decimation transformation for a two-bond
Ising segment.

Consider the hierarchical lattice in Fig.~\ref{lfig1}(a).  If ${\cal
P}_\text{init}$ is the initial probability distribution, a series of
pairwise convolutions which yields the total convolution of
Eq.~\eqref{eq:3} for this lattice is: (i) a bond-moving convolution
of ${\cal P}_\text{init}$ with itself, yielding ${\cal P}_1$; (ii) a
bond-moving convolution of ${\cal P}_1$ with ${\cal P}_\text{init}$,
yielding ${\cal P}_2$; (iii) a decimation convolution of ${\cal
P}_2$ with itself, yielding ${\cal P}_3$; (iv) a decimation
convolution of ${\cal P}_3$ with ${\cal P}_2$, yielding ${\cal
P}_\text{final}$.  For the lattice in Fig.~\ref{lfig2}(a), the
series is: (i) a decimation convolution of ${\cal P}_\text{init}$
with itself, yielding ${\cal P}_1$; (ii) a decimation convolution of
${\cal P}_1$ with ${\cal P}_\text{init}$, yielding ${\cal P}_2$;
(iii) a bond-moving convolution of ${\cal P}_2$ with itself,
yielding ${\cal P}_3$; (iv) a bond-moving convolution of ${\cal
P}_3$ with itself, yielding ${\cal P}_4$; (v) a bond-moving
convolution of ${\cal P}_4$ with itself, yielding ${\cal P}_5$; (vi)
a bond-moving convolution of ${\cal P}_5$ with ${\cal P}_2$,
yielding ${\cal P}_\text{final}$.  For the lattice in
Fig.~\ref{lfig3}(a), the series is: (i) a bond-moving convolution of
${\cal P}_\text{init}$ with itself, yielding ${\cal P}_1$; (ii) a
decimation convolution of ${\cal P}_1$ with itself, yielding ${\cal
P}_2$; (iii) a decimation convolution of ${\cal P}_2$ with ${\cal
P}_1$, yielding ${\cal P}_3$; (iv) a bond-moving convolution of
${\cal P}_3$ with itself, yielding ${\cal P}_4$; (v) a bond-moving
convolution of ${\cal P}_4$ with itself, yielding ${\cal P}_5$; (vi)
a decimation convolution of ${\cal P}_\text{init}$ with itself,
yielding ${\cal P}_6$; (vii) a decimation convolution of ${\cal
P}_6$ with ${\cal P}_\text{init}$, yielding ${\cal P}_7$; (viii) a
bond-moving convolution of ${\cal P}_7$ with ${\cal P}_5$, yielding
${\cal P}_\text{final}$.  As for the dual lattices in
Figs.~\ref{lfig1}(b), \ref{lfig2}(b), and \ref{lfig3}(b), the series
of pairwise convolutions are identical to their counterparts above,
except that each bond-moving is replaced by a decimation, and vice
versa.

Since the number of histograms that constitute the probability
distribution increases rapidly with each renormalization iteration,
a binning procedure is used when the desired (large, namely up to
$2.5 \times 10^9$) number of histograms is reached:  Before every
pairwise convolution, the histograms are placed on a grid, and all
histograms falling into the same grid cell are combined into a
single histogram in such a way that the average and the standard
deviation of the probability distribution are preserved. Histograms
falling outside the grid, representing a negligible part of the
total probability, are similarly combined into a single histogram.
Any histogram within a small neighborhood of a cell boundary is
proportionately shared between the adjacent cells.  In the current
study, the binning procedure is done separately for $J>0$ and $J<0$.
After the convolution, the original number of histograms is
reattained.

In the current study, 40,000 bins are generally used, representing
the renormalization-group flows of 80,000 variables, requiring the
calculation of 40,000 local renormalization-group transformations at
each renormalization-group iteration.  The numerical results
converge rapidly with increasing bin number.  For maximal accuracy
in determining the exact locations of the multicritical points, in
the immediate vicinity of these points we used at least 1,000,000
histograms, representing the renormalization-group flows of
2,000,000 variables, requiring the calculation of 1,000,000 local
renormalization-group transformations at each renormalization-group
iteration.  It should thus be noted that our analysis is an exact
numerical solution of Ising spin-glasses on hierarchical lattices.

\section{Results}

\begin{figure}
\includegraphics*[scale=1]{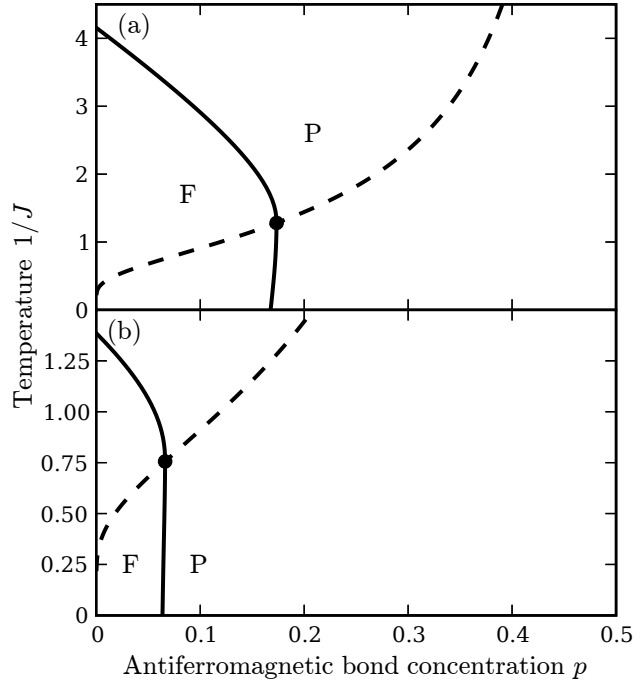}
\caption{Phase diagrams for the two hierarchical lattices in Fig. 1,
with the solid lines indicating second-order phase transitions
between the ferromagnetic (F) and paramagnetic (P) phases.  In each
diagram the multicritical point, separating two different types of
second-order boundary, is marked by a dot, and the Nishimori
symmetry line is drawn dashed.  The phase diagrams were calculated
with 40,000 probability bins, except for the vicinity of the
multicritical points, where for higher precision 1,000,000
probability bins were used.}
\end{figure}

\begin{figure}
\includegraphics*[scale=1]{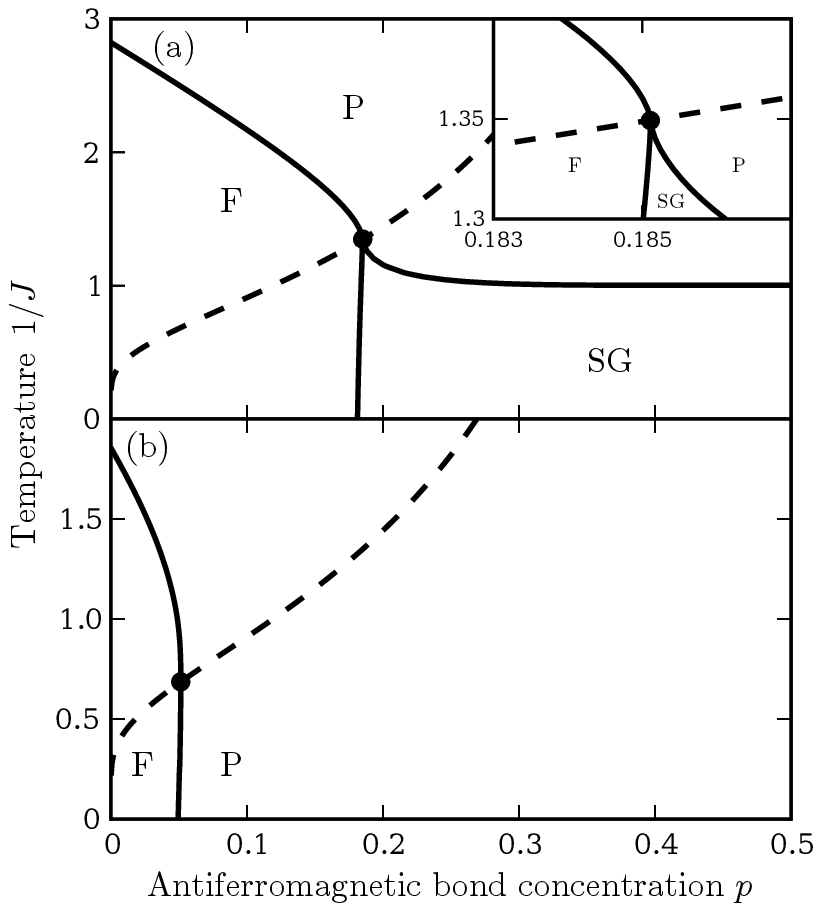}
\caption{Phase diagrams for the two hierarchical lattices in Fig. 2,
with the solid lines indicating second-order phase transitions
between the ferromagnetic (F), paramagnetic (P), and spin-glass (SG)
phases. In each diagram the multicritical point is marked by a dot,
and the Nishimori symmetry line is drawn dashed.  The phase diagrams
were calculated with 40,000 probability bins (250,000 bins for the
inset in the top figure), except for the vicinity of the
multicritical points, where for higher precision $10^6$ and $2.5
\times 10^9$ probability bins were used in (a) and (b)
respectively.}
\end{figure}

\begin{figure}
\includegraphics*[scale=1]{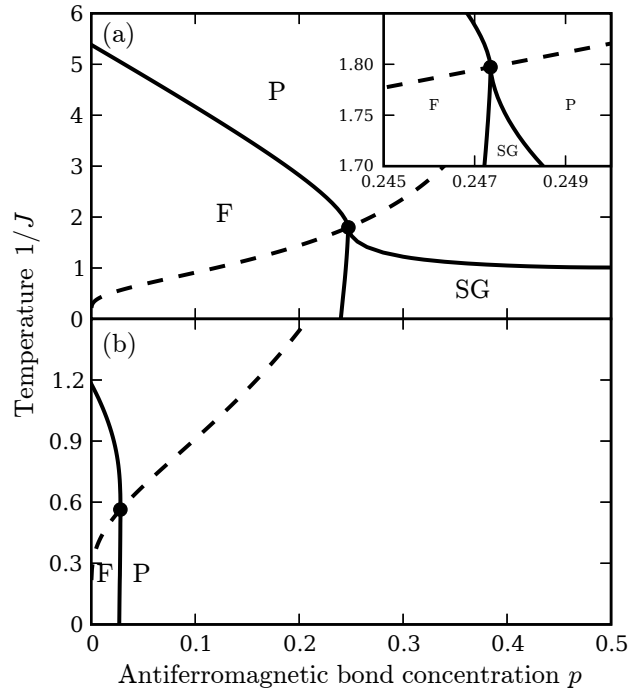}
\caption{Phase diagrams for the two hierarchical lattices in Fig. 3,
with the solid lines indicating second-order phase transitions
between the ferromagnetic (F), paramagnetic (P), and spin-glass (SG)
phases. In each diagram the multicritical point is marked by a dot,
and the Nishimori symmetry line is drawn dashed.  The phase diagrams
were calculated with 40,000 probability bins (250,000 bins for the
inset in the top figure), except for the vicinity of the
multicritical points, where for higher precision $2.5 \times 10^7$
and $2.25 \times 10^8$ probability bins were used in (a) and (b)
respectively.}
\end{figure}

Global phase diagrams for the various hierarchical lattices are
obtained from the renormalization-group flows of the probability
distributions.  Each phase has a corresponding sink, namely a
completely stable fixed distribution.  The boundaries between phases
flow to unstable fixed distributions, analysis of which yields the
order of the phase transition and the values of the critical
exponents of second- and higher-order transitions.  All the phase
diagrams are plotted in terms antiferromagnetic bond concentration
$p$ versus temperature $1/J$.  The diagrams are symmetric around $p
= 1/2$, with the ferromagnetic phase in the $p < 1/2$ half-space
mapping onto the antiferromagnetic phase in the $p > 1/2$
half-space. Thus in the figures only the $p < 1/2$ portions are
shown.

Fig.~4(a) and (b) show the phase diagrams for the dual pair of
hierarchical lattices in Fig.~1(a) and (b) respectively.  The phase
structure of both diagrams is topologically identical to that of the
$d=2$ Ising spin-glass on a square lattice, which is only natural
considering that the $d=2$, $b=3$ Migdal-Kadanoff recursion
relations are exact on these hierarchical
lattices.\cite{BerkerOstlund}  The $p=0$ transition temperatures of
the two models are related by the duality algebra~\cite{Syozi}

\begin{equation}\label{eq:18}
\sinh (2J_{1c})\sinh (2J_{2c}) = 1,
\end{equation}
which is also true for the two other pairs of mutually dual
hierarchical models.  Furthermore, the $p=0$ transition temperatures
in Fig.~4(a) and (b) are related \cite{Kaufman4} by
\begin{equation}\label{eq:19}
J^{-1}_{1c} = b^{d-1}J^{-1}_{2c},
\end{equation}
since the mappings of the interaction constant in the repetition of
renormalization-group transformations differs only by an initial
bond strengthening by a factor of $b^{d-1}$; note that
Eq.~\eqref{eq:19} does not apply to $0<p<1$, since there the
bond-moving is not a mere multiplicative strengthening, but a
$(b^{d-1})$-fold convolution of the probability distributions that
alters this distribution in a non-simple way.  Eq.~\eqref{eq:19} is
also not applicable to the two other pairs of mutually dual models,
since the repetition of renormalization-group transformations are
not differentiated by only a preliminary bond-moving.

In each of Fig.~4(a) and (b), a ferromagnetic phase at low
temperatures and low $p$ is separated from the disordered
paramagnetic phase by two second-order phase boundaries, meeting at
a multicritical point.  (In a narrow neighborhood of all
multicritical points in our results, reentrance is observed:
paramagnetic, then ferromagnetic, then paramagnetic or spin-glass
phases are encountered as temperature is lowered at fixed $p$.)  The two
second-order boundaries flow to distinct unstable probability
distributions with different critical exponents, constituting a
strong violation of universality~\cite{Migliorini} and consistent
with the prediction, generally, of the absence of first-order
transition under quenched randomness in $d=2$.~\cite{Hui}  As
expected from symmetry considerations, the multicritical points
fall~\cite{LeDoussal, LeDoussal2, Hartford} precisely on the
Nishimori line~\cite{Nishimori} as seen in Table I. As also seen in
Table I, $H(p_{1m}) + H(p_{2m}) = 1.0172$, so that the conjecture is
realized to a very good approximation.

\begin{table*}
\begin{tabular}{|c|c|c|c|c|c|c|c|}
\hline \parbox{0.75in}{Figure} &
\parbox{1.2in}{$p_{1m},J^{-1}_{1m}$} &
\parbox{0.5in}{$J^{-1}_{1N}$} & \parbox{1.2in}{$p_{2m}, J^{-1}_{2m}$} &
\parbox{0.5in}{$J^{-1}_{2N}$} &
 \parbox{0.5in}{$H(p_{1m})$} & \parbox{0.5in}{$H(p_{2m})$} & $H(p_{1m}) + H(p_{2m})$ \\
\hline\hline Fig. 4 &  $0.1735,1.2810$ & 1.2810 & $0.06620,0.7557$ &
0.7557 &
0.6656 & 0.3516 & 1.0172 \\
\hline Fig. 5 & $0.1851, 1.3494$ & 1.3494 & ~$0.05128,0.68546$ &
~0.68545 & 0.6911 & 0.2918 & 0.9829 \\
\hline Fig. 6 & $0.2473,1.7973$ & 1.7973 & $0.02796, 0.5636$ &
0.5636 & 0.8070 & 0.1840 & 0.9911 \\
\hline
\end{tabular}
\caption{Locations of the multicritical points in the phase diagrams
of Figs.~4-6 (corresponding to the hierarchical lattices of
Figs.~1-3).  $J^{-1}_{iN}$ is the value calculated from $p_{im}$
using Eq.~\eqref{eq:1j} for the Nishimori line and turns out equal
to $J^{-1}_{im}$, for both $i=1$ and 2.  The quantities $H(p_{im})$
that enter the conjecture and their sums are also given.}
\end{table*}

Fig.~5 shows the phase diagrams for the dual pair of hierarchical
lattices in Fig.~2.  While Fig.~5(b) has the same phase topology as
the diagrams in Fig.~4, being at $d=1.5$ below the spin-glass
lower-critical dimension, a different structure occurs in Fig.~5(a).
Here the $d=3$, $b=3$ Migdal-Kadanoff relations are exact on the
hierarchical lattice, and for low temperatures in the vicinity of
$p=1/2$ there exists a spin-glass phase.  The multicritical point
occurs where the ferromagnetic, paramagnetic, and spin-glass phases
meet.  As expected both multicritical points lie directly on the
Nishimori line.  From Table I we see that $H(p_{1m}) + H(p_{2m}) =
0.9829$, so that the conjecture is realized to a very good
approximation, even when the mutually dual models belong to
different dimensionalities $d$ and have different phase diagram
topologies at the multicritical points of the conjecture.

The phase diagram structures in Fig.~6, corresponding to the dual
pair of hierarchical lattices in Fig.~3, are similar to those of
Fig.~5, illustrating dimensions above and below the spin-glass
lower-critical dimension.  Again the multicritical points for both
cases lie directly on the Nishimori line. In this case $H(p_{1m}) +
H(p_{2m}) = 0.9911$, and the conjecture is realized to a very good
approximation, again for mutually dual models belonging to different
dimensionalities $d$ and having different phase diagram topologies
at the multicritical points of the conjecture.

Thus, we find that for all three mutually dual pairs of hierarchical
lattices, the conjecture relating the locations of the multicritical
points is satisfied to a very good approximation.  This is all the
more remarkable, since, as seen in Table I, the contributions of
$H(p_{1m})$ and $H(p_{2m})$ to the conjecture are strongly
asymmetric.  However, it should be noted that
(1.0172,0.9829,0.9911), while being very close to 1, are different
from integer 1.  In our numerical implementation of the convolutions
of the probability distributions, the results have converged to the
precision of the digits shown in Table I.  Further increase of the
already very large number of probability bins does not change the
entries in the table.  Further tests of the conjecture, using other
systems, would be very useful.  Similar to our current study, the use of 
hierarchical lattices to study phenomena linked to mutually dual lattices, 
e.g., Ref.~\cite{CamiaNewman}, would also be very useful.

\begin{acknowledgments}
We thank H. Nishimori and K. Takeda for useful correspondance.  This
research was supported by the Scientific and Technical Research
Council of Turkey (T\"UBITAK) and by the Academy of Sciences of
Turkey.  MH gratefully acknowledges the hospitality of the Feza
G\"ursey Research Institute and of the Physics Department of
Istanbul Technical University.
\end{acknowledgments}


\newpage

\begin{thebibliography}{}
\bibitem{Nishimori} H. Nishimori, Prog. Theor. Phys. {\bf 66}, 1169 (1981).
\bibitem{NishimoriNemoto} H. Nishimori and K. Nemoto,
  J. Phys. Soc. Japan {\bf 71}, 1198 (2002).
\bibitem{Maillard} J.-M. Maillard, K. Nemoto, and H. Nishimori, J. Phys. A:
  Math. Gen. {\bf 36}, 9799 (2003).
\bibitem{Takeda2} K. Takeda and H. Nishimori, Nucl. Phys. B {\bf 686},
  377 (2004).
\bibitem{Takeda1} K. Takeda, T. Sasamoto, and H. Nishimori,
  J. Phys. A: Math. Gen. {\bf 38}, 3751 (2005).
\bibitem{BerkerOstlund} A.N. Berker and S. Ostlund, J. Phys. C {\bf
  12}, 4961 (1979).
\bibitem{Kaufman} M. Kaufman and R.B. Griffiths, Phys. Rev. B {\bf 24}, 496 (1981).
\bibitem{Kaufman2} M. Kaufman and R.B. Griffiths, Phys. Rev. B {\bf 30}, 244 (1984).
\bibitem{Kaufman3} M. Kaufman and D. Andelman, Phys. Rev. B {\bf 29}, 4010 (1984).
\bibitem{Kaufman4} M. Kaufman, Phys. Rev. B {\bf 30}, 413 (1984).
\bibitem{Itzykson} C. Itzykson and J.M. Luck, Proceedings of the Brasov International Summer School (1984).
\bibitem{Ottavi} H. Ottavi and G. Albinet, J. Phys. A {\bf 20}, 2961 (1987).
\bibitem{LeDoussal} P. Le Doussal and A. Georges, Yale University Report No. YCTP-P1-88
(1988).
\bibitem{LeDoussal2} P. Le Doussal and A.B. Harris, Phys. Rev. Lett. {\bf 61}, 625 (1988).
\bibitem{Hartford} E.J. Hartford and S.R. McKay, J. Appl. Phys. {\bf 70}, 6068 (1991).
\bibitem{McKay} S.R. McKay, A.N. Berker, and S. Kirkpatrick, Phys. Rev. Lett. {\bf 48}, 767
(1982).
\bibitem{Migliorini} G. Migliorini and A.N. Berker, Phys. Rev. B {\bf 57}, 426
(1998).
\bibitem{Andelman} D. Andelman and A.N. Berker, Phys. Rev. B {\bf 29}, 2630
(1984).
\bibitem{Falicov2} A. Falicov, A.N. Berker, and S.R. McKay, Phys. Rev. B
{\bf 51}, 8266 (1995).
\bibitem{Domany} E. Domany, S. Alexander, D. Bensimon, and L.P. Kadanoff, Phys. Rev. B
{\bf 28}, 3110 (1983).
\bibitem{Langlois} J.-M. Langlois, A.-M.S. Tremblay, and B.W. Southern, Phys. Rev. B {\bf 28}, 218 (1983).
\bibitem{Stinchcombe} R.B. Stinchcombe and A.C. Maggs, J. Phys. A {\bf 19}, 1949 (1986).
\bibitem{Haddad} T.A.S. Haddad, S.T.R. Pinho, and S.R. Salinas, Phys. Rev. E {\bf 61}, 3330 (2000).
\bibitem{Le} J.-X. Le and Z.R. Yang, Phys. Rev. E {\bf 69}, 066107 (2004).
\bibitem{daSilveira} R.A. da Silveira and J.-P. Bouchaud, Phys. Rev. Lett. {\bf 93}, 015901 (2004).
\bibitem{Migdal} A.A. Migdal, Zh. Eksp. Teor. Fiz. {\bf 69}, 1457 (1975) [Sov. Phys. JETP {\bf 42}, 743 (1976)].
\bibitem{Kadanoff} L.P. Kadanoff, Ann. Phys. (N.Y.) {\bf 100}, 359 (1976).
\bibitem{ErbasTuncerYucesoyBerker} A. Erba\c{s}, A. Tuncer, B. Y\"{u}cesoy, and A.N. Berker, Phys. Rev. E {\bf 72}, 026129 (2005).
\bibitem{Syozi} I. Syozi, in Phase Transitions and Critical Phenomena, C. Domb and M.S. Green, eds. (Academic, London, 1972),
vol.1, pp.270-329.
\bibitem{Hui} K. Hui and A.N. Berker, Phys. Rev. Lett. {\bf 62}, 2507 (1989).
\bibitem{CamiaNewman}  F. Camia and C.M. Newman, J. Stat. Phys {\bf 114}, 1199 (2004).
\end{thebibliography}
\end{document}